\documentclass[12pt]{iopart}

\usepackage{graphicx}

\usepackage{bm}

\def\beq{\begin{equation}}
\def\enq{\end{equation}}
\def\ba{\begin{eqnarray}}
\def\ea{\end{eqnarray}}

\def\<{<\!\!}
\def\>{\!\!>}
\def\<{\langle}
\def\>{\rangle}

\def\aprge{\buildrel > \over {_{\sim}}}

\begin{document}
\input{epsf}

\title{Ultrahigh-energy neutrino flux as a probe of large 
extra-dimensions}

\author{Joseph Lykken$^{1}$, Olga Mena$^{2}$ and Soebur Razzaque$^{3}$}

\address{$^1$ Theoretical Physics Dept.,
Fermi National Accelerator Laboratory, Batavia, IL 60510, USA}
\address{$^2$ INFN Sez.\ di Roma,
Dipartimento di Fisica, Universit\`{a} di Roma``La Sapienza'', P.le
A.~Moro, 5, I-00185 Roma,Italy}
\address{$^3$Department of Astronomy \& Astrophysics, 
Department of Physics, Pennsylvania State University, University Park,
PA 16802, USA}

\date{\today}

\begin{abstract}
A suppression in the spectrum of ultrahigh-energy (UHE, $\aprge
10^{18}$~eV) neutrinos will be present in extra-dimensional scenarios,
due to enhanced neutrino-antineutrino annihilation processes with the
supernova relic neutrinos.  In the $n> 4$ scenario, being $n$ the
number of extra dimensions, neutrinos can not be responsible for the
highest energy events observed in the UHE cosmic ray spectrum.  A
direct implication of these extra-dimensional interactions would be
the absence of UHE neutrinos in ongoing and future neutrino
telescopes.
\end{abstract}

\submitto{Journal of Cosmology and Astroparticle Physics}

\maketitle

\section{Introduction}

Experimental high-energy neutrino astronomy is developing very
rapidly.  There exist a number of experiments (AMANDA
II~\cite{amanda}, RICE~\cite{rice}, ANITA~\cite{anita},
Icecube~\cite{Icecube}, ANTARES~\cite{Antares}) that are currently
analyzing or starting to take data.  In the future there are planned
projects (ARIANNA~\cite{arianna}, AURA, NEMO, ACORNE) that will
benefit from improved detection techniques and larger effective
detection volumes.

A guaranteed source of UHE neutrino fluxes are the so-called
cosmogenic GZK neutrinos, which are originated by the interactions of
extragalactic UHE cosmic ray (CR) protons with CMB photons dominantly
via $\Delta^{+}$ processes and subsequent charged pion decays.
Cosmogenic neutrinos are typically characterized by a spectrum peaking
in the $10^{17-19}$ eV energy range, depending on the redshift of the
CR sources.  Ongoing and future experiments expect to detect a few GZK
neutrino events; the precise number depends on the full exposure of
the instruments as well as on the production model.  Direct emission
of UHE neutrinos from the CR sources is expected but uncertain. Decays
of topological defects or supermassive particles, leftover fossils
from the GUT era, is speculative.  Nevertheless, both mechanisms would
produce neutrino fluxes with energies comparable to or higher than
those associated to the GZK fluxes.  These neutrinos could interact
with $1.95\,^{\circ}$K CMB neutrinos (C$\nu$B) via the standard model
(SM) reaction $\nu \bar{\nu}\rightarrow Z^{0}$, provided that they are
extremely energetic
($10^{22-25}$~eV)~\cite{weiler,roulet,weiler2,quigg}. We do not
explore these speculative neutrino fluxes in the present study.

In this study, we focus on the depletion of the GZK cosmogenic
neutrino fluxes via strongly interacting annihilation processes with
\emph{other neutrino relics} that also permeate the universe: the
diffuse supernova relic neutrinos (DSN$\nu$), that represent the flux
of neutrinos from all supernova explosions that occurred during the
universe's history.  The DSN$\nu$ direct detection is still elusive.
The most stringent experimental current limit to the DSN relic
$\bar{\nu}_{e}$ flux is $1.2$~cm$^{-2}$s$^{-1}$ at $90\%$ CL, from the
SuperKamiokande experiment~\cite{SK}. The presence of strongly
interacting processes, such as the exchange of massive spin-2
particles in theories of large extra-dimensions~\cite{ah,GRW,joe}, can
modify the $\nu \bar{\nu}$ annihilation cross section.  This effect
would take place at high values of the squared center-of-mass energy
$s$, yielding a $\nu \bar{\nu}$ annihilation cross section that is
larger than the cross section for the SM process $\nu
\bar{\nu}^{SM}\rightarrow Z^{0}$. In principle, the UHE cosmogenic
neutrinos can annihilate with both the C$\nu$B~\cite{Goyal:2000ma} and
DSN$\nu$ via extra-dimensional enhanced cross sections, which we
discuss next.

\section{Neutrino annihilation in extra-dimensional models} 
We consider the following annihilation cross sections for $n$ extra
dimensions \cite{joe,Goyal:2000ma}
\ba
\sigma_{\nu{\bar \nu} \to g_{KK}} &=& (\pi^2/s) (s/M_S^2)^{n/2 +1}
\nonumber \\
\sigma_{\nu{\bar \nu} \to f{\bar f}} &=& (\pi/60s) (s/M_S^2)^{n+2}
{\cal F}^2 \nonumber \\
\sigma_{\nu{\bar \nu} \to \gamma\gamma} &=& 
3\sigma_{\nu{\bar \nu} \to f{\bar f}},
\label{eq:large_x-section}
\ea
respectively to produce KK gravitons, fermion- and $\gamma$- pairs.
Here ${\cal F}^2 = \pi^2 +4 I^2 (M_S/\sqrt{s})$ and we use
$I(M_S/\sqrt{s})$ as given in Ref.~\cite{joe}.  The ``new physics''
scale $M_S$ is constrained from astrophysical considerations such as
star cooling by graviton emission~\cite{ah,astro-limit} and from
collider searches~\cite{d0}.  In particular, we use $M_S = 701$~TeV,
25.5~TeV and 2.77~TeV for $n=2$, 3 and 4, the most stringent current
constraints from heating of
neutron-stars~\cite{astro-limits-latest}. For $n>4$, the most
stringent lower bounds are from the D0 collider experiment at the
Tevatron, which sets the $95\%$~CL limits for n=5, 6 and 7 equal to
0.97~TeV, 0.9~TeV and 0.85~TeV, respectively~\cite{d0}.  In the $n=5$
scenario, the total $\nu \bar{\nu}$ annihilation cross section is
$\simeq 4 \times 10^{-19}$~cm$^{2}$ at $\sqrt{s}\simeq 14$~TeV, which
roughly corresponds to a $10^{19}$~eV GZK neutrino interacting with a
$10$~MeV DSN relic antineutrino.  The cross section quoted above is
therefore many orders of magnitude larger than the SM cross section
$\sigma^{SM}_{\nu \bar{\nu} \rightarrow all} \simeq 8\times
10^{-34}$~cm$^{2}$ at the same $\sqrt{s}$ and scales as $\sim s^6$ for
$n=5$ and $s
\gg M_S$.

The neutrino interactions in Eqs.~(\ref{eq:large_x-section}) are
independent of the neutrino flavor.  Brane-bulk couplings are flavor
blind and consequently the exchange of the KK gravitons is unaffected
by the electron, muon or tau nature of the DSN (anti)neutrinos, except
corrections proportional to the squared mass splittings divided by
$s$, which are negligible ($\mathcal{O}(10^{-27})$)~\footnote{This
flavor blindness character of the extra-dimensional model presented
here no longer holds if one or more of the neutrino species are in the
bulk. Such a possibility is not considered through the present
discussion.}.

A word of caution is needed here regarding the extra-dimensional
scenario, which is an effective theory valid for $s \sim M^2_{S}$.  At
some energy scale $s \sim M^2_{S}$, this theory is supposed to match
onto a more fundamental theory of quantum gravity.  It is not known
how to do this matching. A phenomenological approach is to assume that
the neutrino interaction cross sections in the $s \sim M^2_{S}$ energy
range behave similarly to the cross sections in the $s
\sim M^2_S$ energy regime, up to some cutoff $\Lambda$.
The value of $\Lambda$ is presumably somewhere between $M_S$
and $E_{\rm max}$, where the latter is the scale at which
perturbative unitarity would be violated~\cite{GRW}. For the
models we consider $E_{\rm max}$ is always greater than
$5.6 M_S$.

Within the context of extra-dimensional models, the $\nu N$ cross
sections will be enhanced as
well~\cite{anchordoqui,emparan,anchordoqui2,olinto,ina,dan,ringwald},
providing a possible explanation for the events above the GZK cut-off
as explored in Refs.~\cite{domokos,Goyal:2000ma,panda,jain}. However,
as we will discuss shortly, $10^{20}$~eV neutrinos would annihilate
with DSN$\nu$ on their flight to the Earth rather than producing an
extended air shower in the atmosphere, via enhanced $\nu N$ cross
section, in the large extra-dimensional models. The advantage of
exploring the $\nu\bar{\nu}$ annihilation channel is that
extradimensional signatures would occur at lower energy, compared to
the signatures in the commonly explored $\nu N$ interaction.

\section{Supernova relic neutrino density and UHE neutrino
propagation} 

A number of authors have predicted the DSN$\nu$ flux. For a recent
appraisal of the theoretical and computational status, see
Ref.~\cite{dsnb} and references therein.  Here we follow closely the
derivation given in Ref.~\cite{Ando:2004hc}.  A fit to the neutrino
spectra from numerical simulations of a SN
is~\cite{Keil:2002in,Totani:1997vj}
\ba
\frac{dN_\nu^0}{dE_\nu} =
\frac{(1+\beta_\nu)^{1+\beta_\nu} L_{\nu}}{\Gamma(1+\beta_\nu)  
{\bar E}_{\nu}^2}
\left( \frac{E_\nu}{{\bar E}_\nu} \right)^{\beta_\nu}
e^{-(1+\beta_\nu)E_{\nu}/{\bar E}_\nu},
\ea
where the average energy ${\bar E}_\nu = 15.4$~MeV and $21.6$~MeV
respectively for ${\bar \nu}_e$ and $\nu_x$ corresponding to all other
non-electron anti-neutrino and neutrino flavors.  The spectral indices
are $\beta_{{\bar \nu}_e} = 3.8$ and $\beta_{\nu_x} = 1.8$ while the
total neutrino energies are $L_{{\bar \nu}_e} \simeq L_{\nu_x} =
5\times 10^{52}$~erg.  For $\nu_e$, we use ${\bar E}_{\nu_e} = 11$~MeV
\cite{Totani:1997vj}, $L_{\nu_e} \simeq L_{{\bar \nu}_e}$ and
$\beta_{\nu_e} = \beta_{{\bar \nu}_e}$.  Neutrino conversion inside
the star mixes the different neutrino flavors and therefore the relic
(anti) neutrino flavor spectra at the stellar surface will differ from
the original ones.  The final flavor spectra will depend on the
neutrino mass ordering (normal versus inverted) and the adiabaticity
of the transitions in the resonance layers, see
Ref.~\cite{Dighe:1999bi} for a complete description.  As we will
explain further below, the $\nu \bar{\nu}$ interactions we explore
here are flavor blind and therefore the GZK (anti) neutrino will
interact with the three (neutrino) antineutrino flavors. Therefore we
do not need to account for conversion effects and the relevant
quantity would be the total antineutrino (neutrino) SN relic neutrino
spectra, given by:
\ba
\frac{dN_{{\bar \nu} (\nu)}}{dE_\nu} = 
\frac{dN_{{\bar \nu}_e (\nu_e)}^0}{dE_\nu} + 
2 \frac{dN_{\nu_x}^0}{dE_\nu},
\ea
that is, the sum of the three flavor spectra.

The redshift-dependent SN rate is a fraction $0.0122 M_\odot^{-1}$ of
the star formation rate and is given, e.g. SF1 model in
Ref.~\cite{sfr}, by
\ba
R_{sn}(z) &=& 0.0122\times 0.32 h_{70} 
\frac{\exp(3.4z)}{\exp(3.8z)+45} 
\nonumber \\ && \times 
\left[ \frac{\Omega_m (1+z)^3 + \Omega_\Lambda}{(1+z)^{3}}
\right]^{1/2} ~{\rm yr}^{-1}~{\rm Mpc}^{-3}
\ea
with a Hubble constant $H_0 = 70h_{70}$~km~s$^{-1}$~Mpc$^{-1}$ and
$\Lambda$CDM cosmology.  The other parameters are $\Omega_m =0.3$ and
$\Omega_\Lambda = 0.7$.  The differential number density of SN relic
neutrinos at present from all past SNe up to a maximum redshift
$z_{\rm sn,max}$ is then~\cite{Ando:2004hc}
\ba
\frac{dn_{{\bar \nu} (\nu)}}{dE_{\nu}} = \int_0^{z_{\rm sn,max}} dz
\frac{dt}{dz} (1+z) R_{sn}(z)
\frac{dN_{{\bar \nu} (\nu)}}{dE'_{\nu}}.
\label{snrn_density}
\ea
Here $(dt/dz)^{-1} = -H_0 (1+z) [\Omega_m (1+z)^3 + \Omega_\Lambda
]^{1/2}$ and $E_{\nu} = E'_{\nu}/(1+z)$ is the redshift-corrected
observed energy.

While the number density of the DSN$\nu$ ($10^{-9}$~cm$^{-3}$ for the
sum of the three (anti)neutrino flavors) is orders of magnitude
smaller than those for the C$\nu$B relics ($56$~cm$^{-3}$ per each
(anti)neutrino flavor), the average energy of the DSN$\nu$ is tens of
MeV, compared to the $10^{-4}$ eV for C$\nu$B relics.  Therefore, the
UHE neutrino mean-free-path, $mfp=1/\sigma_{\nu {\bar
\nu}} n_\nu$ is many orders of magnitude smaller in the
case of the less abundant, but more energetic DSN$\nu$ compared to the
C$\nu$B relics. If the strongly interacting processes deplete the UHE
cosmogenic neutrino fluxes, the dominant attenuator will be the
DSN$\nu$ targets, which we discuss more quantitatively below.

An UHE $\nu$ of observed energy $E_{\nu,\rm uhe}$ may interact with a
DSN$\nu$ at redshift $z'$ on its way via processes in
Eq.~(\ref{eq:large_x-section}) and annihilate.  The corresponding
$s\simeq 2 E_{\nu,\rm uhe} (1+z') E_{\nu,\rm sn} (1+z)$, ignoring the
$\nu$ masses.  We use the maximum SN $\nu$ energy to be $E'_{\nu,\rm
sn,max} = 60$~MeV in the SN rest frame. The inverse {\em mfp} for
$\nu{\bar \nu}$ annihilation is then
\ba
{\mathcal L}^{-1} (E_{\nu,\rm uhe};z') = 
\int_{z'}^{z_{\rm sn,max}} dz \frac{dt}{dz} (1+z) R_{sn}(z) \nonumber \\
\times \int_0^{E'_{\nu,\rm sn,max}}
dE'_{\nu,\rm sn} \frac{dN_{{\bar \nu},\rm sn}}{dE_{\nu,\rm
sn}} \sigma_{\nu {\bar \nu}} (s).
\label{nunu_mfp}
\ea
The {\em mfp} for a $10^{19}$~eV neutrino to annihilate with a
DSN$\nu$ via the SM process $\bar{\nu}^{SM}_{\nu{\bar\nu}\rightarrow
all}$ is $10^{18}$~Mpc, which exceeds the Hubble distance.  Within the
$n=5$ extra-dimensional model, the annihilation cross section is
greatly enhanced at high energies, and the {\em mfp} for a
$10^{19}$~eV neutrino is $\sim 12$~Mpc in our local universe ($z'\sim
0$), which is less than the GZK radius.  Even for the $n=4$
extra-dimensional model, the {\em mfp} for the highest energy CR,
$3\times 10^{20}$~eV, is $\sim 127$~Mpc which is comparable to the GZK
radius.  To explain GZK CR data with UHE neutrinos through enhanced
$\nu N$ cross section requires $n> 4$.  Thus UHE neutrinos propagating
from outside the GZK radius can not be the candidates for GZK CR
events, since they would be absorbed by DSN$\nu$.

We can now calculate the survival probability for an UHE $\nu$ created
at redshift $z_{\rm uhe}$ to reach Earth as
\ba
P(E_{\nu,\rm uhe}; z_{\rm uhe}) = \exp \left[ 
-c\int_0^{z_{\rm uhe}} dz' 
\frac{dt}{dz'} {\mathcal L}^{-1} (E_{\nu,\rm uhe}; z') \right] 
\nonumber \\
= \exp \left[ -{\mathcal K} \frac{c}{H_0^2}
\int_0^{z_{\rm uhe}} \frac{dz'}
{(1+z')\sqrt{\Omega_m (1+z')^3 + \Omega_\Lambda}} \right. \nonumber \\
\times \left. \int_{z'}^{z_{\rm sn,max}} \frac{dz}{(1+z)^{3/2}}
\frac{\exp(3.4z)}{\exp(3.8z)+45} \right. \nonumber \\
\times \left. \int_0^{E'_{\nu,\rm sn, max}}
dE_{\nu,\rm sn} \frac{dN_{{\bar \nu},\rm sn}}{dE_{\nu,\rm sn}}  
\sigma_{\nu {\bar \nu}} (s) \right],
\label{nunu_attenuation}
\ea
where ${\mathcal K} c/H_0^2 \approx 2.45\times 10^{-38}
h_{70}^{-1}$~cm$^{-2}$ and the differential SN $\nu$ spectrum is
$dN_{{\bar \nu},\rm sn}/dE_{\nu,\rm sn} \approx 10^{49}$~MeV$^{-1}$.
Large $\nu{\bar\nu}$ cross section then suppresses UHE neutrinos. We
discuss UHE $\nu$ fluxes that will be attenuated by $\nu{\bar
\nu}$ annihilation next.

\section{Ultrahigh-energy neutrino flux} The CR energy generation rate
per unit volume in our local universe in the energy range
$10^{19-21}$~eV is $P_{\rm CR} \approx 5\times
10^{44}$~erg~Mpc$^{-3}$~yr$^{-1}$ \cite{Waxman:1998yy}.  Assuming an
injection spectrum for CR protons $dN_{p}/dE^0_{p} \propto
E_{p}^{-2}$, as typically expected, we define a convenient conversion
formula
\ba
{\mathcal N}_{\rm CR} &=& 
\frac{c}{4\pi H_0} \frac{P_{\rm CR}}{{\rm
ln}(10^{21}/10^{19})} \nonumber \\
&\approx & 7.1\times 10^{-8} h_{70}^{-1} 
~{\rm GeV}~{\rm cm}^{-2}~{\rm s}^{-1}~{\rm sr}^{-1},
\label{cr_norm}
\ea
which is proportional to the CR flux $E_p^2 J_p$ above $10^{19}$~eV.
We will use Eq.~(\ref{cr_norm}) to fix the normalization of UHE $\nu$
fluxes.  The CR sources may also evolve with redshift as $S(z) =
(1+z)^3$ for $z<1.9$, $(1+1.9)^3$ for $1.9<z<2.7$ and
$\exp[(2.7-z)/2.7]$ for $z>2.7$ \cite{Waxman:1998yy}.

The Waxman-Bahcall (WB) bound on UHE $\nu$ flux \cite{Waxman:1998yy}
is based on CRs that interact at their sources and lose all their
energy equally to charged and neutral pions.  The resulting $\nu_\mu$
flux is given by
\ba
E_\nu^2 J_{\nu,\rm WB} = \frac{{\mathcal N}_{\rm CR}}{8}
\int_0^{z_{\rm max}} dz_{\rm uhe} 
\frac{S(z_{\rm uhe}) P(E_\nu; z_{\rm uhe})}
{\sqrt{\Omega_m (1+z_{\rm uhe})^3 + \Omega_\Lambda}}
\label{wb_flux}
\ea
after integrating over CR source evolution and $\nu {\bar \nu}$
annihilation probability in Eq.~(\ref{nunu_attenuation}).

If UHE CRs interact with CMB photons in the local universe then the
resulting GZK neutrino flux would be
\ba
E_\nu J_\nu (z\sim 0) \propto {\mathcal N}_{\rm CR} \int dE_p^0
\frac{dN_p}{dE_p^0} Y(E_p^0, E_\nu, z\sim 0)
\ea
Here $Y$ is called the neutrino yield function as in
Ref.~\cite{Engel:2001hd} and is the number of secondary neutrinos
generated per unit energy interval by a CR proton of energy
$E_p^0$. We use a fit to $Y(E_p^0, E_\nu, z\sim 0)$ corresponding to
$\nu_\mu$ and ${\bar \nu}_\mu$ from a CR proton propagating 200~Mpc as
generated by the SOPHIA Monte Carlo code as reported in
Ref.~\cite{Engel:2001hd}. The GZK $\nu$ spectra are fully evolved by
200~Mpc in our local universe and over smaller distance at higher
redshift. Our calculation shows that this distance is much shorter
than the {\em mfp} for $\nu N$ interactions of UHE CRs with DSN$\nu$
in $n\ge 4$ large extra-dimensional models. Thus we calculate the
effect of $\nu{\bar \nu}$ annihilation assuming that a fully evolved
GZK $\nu$ flux exist at a given redshift of interaction.

The GZK $\nu$ flux integrated over all CR sources, after taking into
account the redshift evolution of the neutrino yield function
$Y(E_p^0, E_\nu, z) = Y(E_p^0(1+z), E_\nu (1+z)^2, z\sim 0)$
\cite{Engel:2001hd}, the source evolution $S(z)$ and finally the
survival probability $P(E_\nu; z_{\rm uhe})$ in
Eq.~(\ref{nunu_attenuation}), is given by
\ba
E_\nu J_{\nu,\rm GZK} &=& {\mathcal N}_{\rm CR}
\int_0^{z_{\rm max}} dz_{\rm uhe} 
\frac{S(z_{\rm uhe}) P(E_\nu; z_{\rm uhe})}
{\sqrt{\Omega_m (1+z_{\rm uhe})^3 + \Omega_\Lambda}}
\nonumber \\ && \times
\int dE_p^s \frac{dN_p}{dE_p^s} Y(E_p^s, E_\nu, z_{\rm uhe}).
\label{gzk_flux}
\ea
In case of no $\nu {\bar \nu}$ annihilation, $P(E_\nu; z_{\rm uhe})=1$
and the flux is the same as in Ref.~\cite{Engel:2001hd}.

\begin{figure}
\centerline {\epsfxsize=3.4in \epsfbox{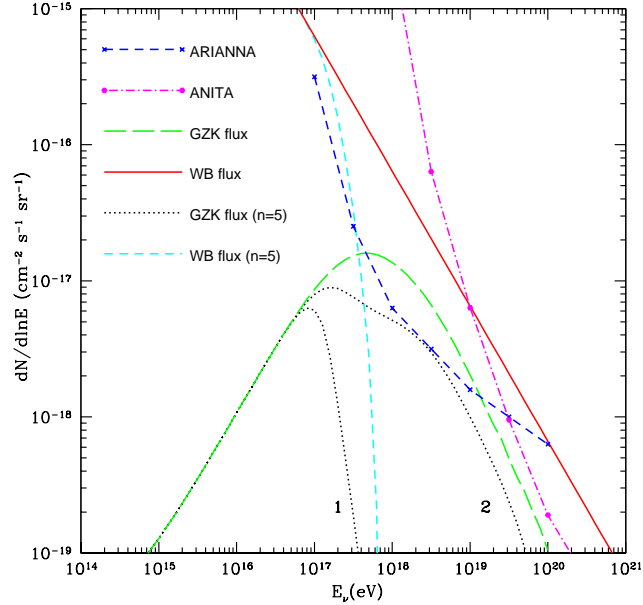}} 
\caption{
UHE $\nu$ fluxes from the cosmic ray protons interacting at the source
(WB) and in CMB (GZK).  If $\nu {\bar \nu}$ annihilation is important,
as in large extra-dimensional models (shown here for $n=5$ case with
dotted curves labeled 1 and 2), then UHE $\nu$ fluxes would be
suppressed.  Also shown are the projected sensitivities for the ANITA
(50 days) and the proposed ARIANNA (6 months) UHE neutrino experiments
at the South Pole.}
\label{fig:flux}
\end{figure}

We have numerically evaluated the GZK flux, both without and with $\nu
{\bar \nu}$ annihilation, using $z_{\rm max} = z_{\rm uhe} = z_{\rm
sn,max} = 5$ and in the energy range $10^{19}~{\rm eV} < E_p^0 <
10^{22}~{\rm eV}$ with an exponential cutoff of the $\propto E_p^{-2}$
spectrum at $3\times 10^{21}$~eV as in Ref.~\cite{Engel:2001hd}.  The
results for the GKZ cosmogenic $\nu_\mu$ flux are depicted in
Fig.~\ref{fig:flux}, assuming a $n=5$ extra-dimensional scenario (the
dotted curves labeled 1 and 2 corresponds to $\Lambda = \infty$ and
20~TeV respectively; allowing the cross sections in
Eq.~(\ref{eq:large_x-section}) to grow below $\sqrt{s} = \Lambda$ and
become flat above).  Also shown is the WB flux without and with $\nu
{\bar \nu}$ annihilation. Notice that the $n=5$ extra-dimensional
scenario leaves a clear imprint on the GZK cosmogenic neutrino fluxes,
which would be abruptly truncated above $E\aprge 10^{17}$~eV. This
characteristic feature in the GZK cosmogenic fluxes could be
recognized by the presence of a \emph{dip} in the neutrino spectra,
provided the detection technique has a low enough energy
threshold. For ongoing and future UHE neutrino experiments with higher
energy thresholds ($E \aprge 10^{17}$~eV), such as ANITA and ARIANNA
shown in Fig.~\ref{fig:flux}, there would be an absence of neutrino
induced events caused by strongly interacting, KK-modes mediated $\nu
\bar{\nu}$ processes.  For the $n < 5$ extradimensional
models, the UHE neutrino flux suppression would occur at UHE neutrino
energies $E \aprge 10^{19-20}$ eV, where the cosmogenic neutrino
fluxes are smaller and consequently, also the statistics expected in
ongoing and future UHE neutrino observatories would be
reduced. 

\begin{figure}
\centerline {\epsfxsize=3.4in \epsfbox{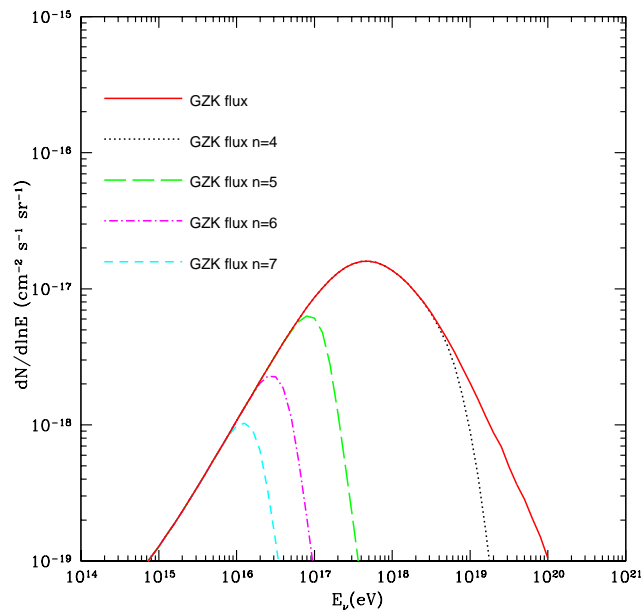}} 
\caption{
The solid line depicts the UHE $\nu$ fluxes from the cosmic ray
protons interacting with CMB photons (GZK neutrino fluxes). The
dotted, long-dashed, dot-dashed and short-dashed curves illustrate the
GZK neutrino fluxes for the case of $n=4$, 5, 6 and 7 extradimensions
exist in nature, respectively.}
\label{fig:flux2}
\end{figure}

\begin{figure}
\centerline {\epsfxsize=3.4in \epsfbox{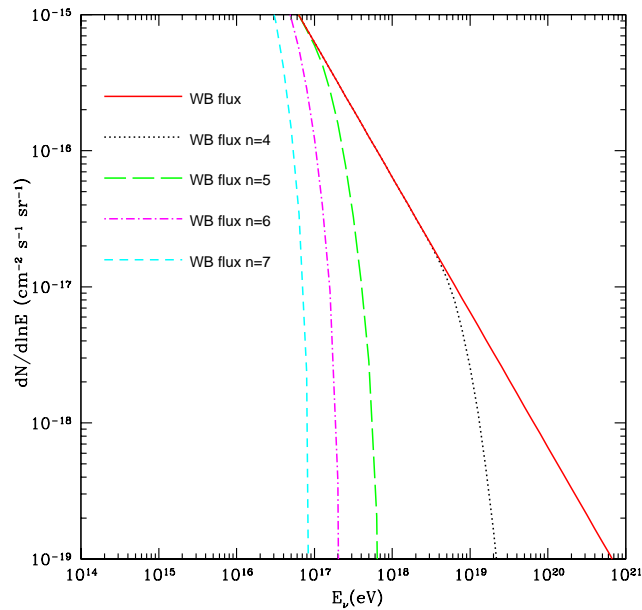}} 
\caption{ 
Same as Fig.~\ref{fig:flux2} but for UHE $\nu$ fluxes from the cosmic
ray protons interacting at the source (WB neutrino fluxes).}
\label{fig:flux3}
\end{figure}

Figure~\ref{fig:flux2} depicts the the GZK cosmogenic $\nu_\mu$ flux
with and without extradimensional suppression for the case nature has
$n=4$, 5, 6 and 7 extra dimensions.  For $n < 4$, the UHE neutrino
flux suppression is subtle and therefore it would be highly
challenging and difficult to detect experimentally.
Figure~\ref{fig:flux3} illustrates the WB flux without and with $\nu
{\bar \nu}$ annihilation. If $n < 5$, tracking the extra-dimensional
induced suppression \emph{dip} would be more difficult in general.
Note that an increase of $\nu N$ cross section, expected in this
scenario, do not significantly increase the detector sensitivity
because of a steeply falling $\nu$~flux and a decreasing angular
acceptance with increasing energy (see, e.g.,~\cite{rice}).

\section{Summary and conclusions} 
We have shown that UHE neutrinos will be absorbed, in theoretical
models that predict fast-rising cross sections such as large
extra-dimensional models, by a diffuse background of 10~MeV neutrinos
provided by all core-collapse SNe in the history of the universe.
Detection of neutrinos from the SN~1987A proves the existence of such
neutrinos, and upcoming megaton detectors will measure the diffuse
flux to a good accuracy.

If there exist $n\ge 5$ large extra-dimensions in nature, and the
DSN$\nu$ flux is detected at the level of the current theoretical
models, then UHE neutrinos can not be the primaries of the super GZK
events, since the UHE neutrino fluxes will suffer a \emph{cutoff} in
their energy spectra in the $10^{16-18}$~eV energy range.  On the
other hand, a detection of GZK neutrinos at energies $E \aprge
10^{18}$~eV could imply the absence of $n \ge 5$ large
extra-dimensions in nature, and therefore eliminating such models. For
$n< 5$ extradimensions, neutrinos could be the UHE CR primaries if the
$\nu N$ cross-section is sufficiently enhanced to mimic hadronic
cross-section.

In case the DSN$\nu$ flux is detected at a much lower level, then the
{\em dip} in the UHE neutrino spectrum, due to absorption by DSN$\nu$,
would be shifted to higher energy. Note that $\nu{\bar\nu}$
annihilation by UHE neutrinos would not produce $\gamma$-rays over the
EGRET limit, since the primary UHE CR interactions with CMB and
infrared photons can not account for the observed diffuse $\gamma$-ray
flux~\cite{gamma}.  Also the GZK CRs are not affected due to large
$\nu N$ cross section, since they are expected to be produced within
$\sim 50$~Mpc, a radius smaller than the $\nu N$ {\em mfp} with
enhanced cross section.

Measuring an enhancement of UHE neutrino cross sections at ongoing or
future neutrino observatories, will be therefore extremely difficult,
since in these scenarios the GZK cosmogenic neutrino fluxes would be
depleted in their way to the Earth via annihilation with the DSN$\nu$
background.

\ack
It is pleasure to thank E.~J.~Ahn, R.~Engel, D.~McKay and S.~Panda for
useful comments.  We also thank an anonymous referee for pointing out
the most stringest current limits on the new physics scale in
literature.  Work supported in parts through NSF grant AST 0307376
(SR) and the European Programme ``The Quest for Unification'' contract
MRTN-CT-2004-503369 (OM).  Fermilab is operated by the Fermi Research
Alliance LLC under contact DE-AC02-07CH11359 with the U.S.\ Dept.\ of
Energy.

\end{document}